\documentclass[12pt]{article}
\usepackage[nosort]{cite}
\usepackage[usenames, dvipsnames]{xcolor}
\usepackage{graphicx}
\usepackage{multicol}
\usepackage{amsfonts}
\usepackage{amssymb}
\usepackage{amsmath}
\usepackage{afterpage}
\usepackage{setspace}
\usepackage{verbatim}
\usepackage{color}
\usepackage{longtable}
\usepackage{caption}
\usepackage{heck}
\usepackage{float}
\usepackage{slashed}
\usepackage[utf8]{inputenc}
\usepackage{braket}
\usepackage{subcaption}
\usepackage{epsfig}
\usepackage{enumerate}
\usepackage{epstopdf}
\usepackage[enableskew, vcentermath]{youngtab}
\usepackage{adjustbox}
\usepackage{multirow}
\usepackage{tikz}
\usepackage[margin=1in]{geometry}
\usepackage{titletoc}
\usepackage{hyperref}
\usepackage[percent]{overpic}
\usepackage{gensymb}
\usepackage{mathtools}
\usepackage{tikz-cd}%
\usepackage[utf8]{inputenc}
\usepackage{epsf}
\usepackage{colortbl}
\usepackage{amsmath}
\usepackage{amsfonts}
\usepackage{amssymb}
\usepackage{graphicx}
\usepackage{color}
\usepackage{psfrag}
\usepackage{cite}
\usepackage{hyperref}
\usepackage{mathrsfs}
\usepackage{tikz}
\usepackage{tikz-feynman}
\usepackage{slashed}
\usepackage{ifpdf}
\usepackage{mdframed}
\usepackage[T1]{fontenc}
\mdfdefinestyle{statementstyle}{
   frametitlerule = true,
   frametitlefont = {\normalfont\bfseries},
   frametitlebackgroundcolor = gray!10,
}
\mdtheorem[style=statementstyle]{statement}{Definition}
\providecommand{\U}[1]{\protect\rule{.1in}{.1in}}
\pdfoutput=1
\newsavebox{\mysavebox}

\hypersetup{colorlinks,citecolor=black,filecolor=black,linkcolor=black,urlcolor=black}
\usetikzlibrary{decorations.markings}

\numberwithin{equation}{section}

\usetikzlibrary{chains}
\allowdisplaybreaks
\tikzset{node distance=2em, ch/.style={circle,draw,on chain,inner sep=2pt},chj/.style={ch,join},every path/.style={shorten >=4pt,shorten <=4pt},line width=1pt,baseline=-1ex}

\newcommand{\ba}{\begin{eqnarray}}
\newcommand{\ea}{\end{eqnarray}}

\newcommand{\be}{\begin{equation}}
\newcommand{\ee}{\end{equation}}

\tikzstyle{startstop} = [rectangle, rounded corners, minimum width=3cm, minimum height=1cm,text centered, draw=black, fill=blue!10]
\tikzstyle{startstop} = [rectangle, rounded corners, minimum width=3cm, minimum height=1cm,text centered, draw=black, fill=blue!10]
\tikzstyle{io} = [trapezium, trapezium left angle=70, trapezium right angle=110, minimum width=3cm, minimum height=1cm, text centered, draw=black, fill=blue!30]
\tikzstyle{process} = [rectangle, minimum width=3cm, minimum height=1cm, text centered, draw=black, fill=orange!30]
\tikzstyle{decision} = [diamond, minimum width=3cm, minimum height=1cm, text centered, draw=black, fill=green!30]
\tikzstyle{arrow} = [thick,->,>=stealth]
\tikzset{->-/.style={decoration={
  markings,
  mark=at position #1 with {\arrow[scale=2.4]{>}}},postaction={decorate}}}
\makeatletter \@addtoreset{equation}{section} \makeatother

\renewcommand{\[}{\left[}

\colorlet{darkblue}{blue!70!black}
\colorlet{darkgreen}{green!70!black}

\hypersetup{
colorlinks=true,
linkcolor=blue,
citecolor=magenta,
}

\begin{document}

\preprint{LMU-ASC 16/23\\ IFT-UAM/CSIC-23-50}

\date{May 2023}

\title{R7-Branes as Charge Conjugation Operators}

\institution{LMU}{\centerline{$^{1}$Arnold Sommerfeld Center for Theoretical Physics, LMU, Munich, 80333, Germany}}
\institution{PENN}{\centerline{$^{2}$Department of Physics and Astronomy, University of Pennsylvania, Philadelphia, PA 19104, USA}}
\institution{PENNmath}{\centerline{$^{3}$Department of Mathematics, University of Pennsylvania, Philadelphia, PA 19104, USA}}
\institution{HARVARD}{\centerline{$^{4}$Instituto de Física Teórica UAM-CSIC, c/Blas Cabrera 13-15, 28049 Madrid, Spain}}

\authors{
Markus Dierigl\worksat{\LMU}\footnote{e-mail: \texttt{m.dierigl@lmu.de}},
Jonathan J. Heckman\worksat{\PENN,\PENNmath}\footnote{e-mail: \texttt{jheckman@sas.upenn.edu}},\\[4mm]
Miguel Montero\worksat{\HARVARD}\footnote{e-mail: \texttt{miguel.montero@csic.es}}, and
Ethan Torres\worksat{\PENN}\footnote{e-mail: \texttt{emtorres@sas.upenn.edu}}
}

\abstract{R7-branes are a class of recently discovered non-supersymmetric real codimension-two
duality defects in type IIB string theory
predicted by the Swampland Cobordism Conjecture.
For type IIB realizations of 6D SCFTs with $\mathcal{N} = (2,0)$ supersymmetry, wrapping an R7-brane
``at infinity'' leads to a topological operator associated with a zero-form charge conjugation symmetry that squares to the identity.
Similar considerations hold for those theories obtained from further toroidal compactification, but this can be obstructed by bundle curvature effects. Using some minimal data on the topological sector of the R7-branes, we extract the associated fusion rules for these charge conjugation operators. More broadly, we sketch a top down realization of various topological operators /  interfaces associated
with $\mathsf{C}$, $\mathsf{R}$, and $\mathsf{T}$ transformations. We also use holography to provide strong evidence for the existence of the R7-brane which is complementary to the Cobordism Conjecture. Similar considerations apply to other string-realized QFTs with symmetry operators constructed via non-supersymmetric branes which carry a conserved charge.}

\maketitle

\setcounter{tocdepth}{2}


\newpage

\section{Introduction}\label{sec:INTRO}

Symmetries provide a powerful organizing tool in the study of quantum fields
and gravity. Recently, it has been shown that the structures of symmetries in
physical systems are intimately tied with topological structures. In the context of quantum
field theory (QFT), such generalized symmetries provide a framework for
understanding many of these features \cite{Gaiotto:2014kfa}, and this has by now led to a number
of new developments both in the study of higher-form, higher-group, as well as
non-invertible / categorical generalizations.\footnote{See e.g.,
\cite{Gaiotto:2014kfa, Gaiotto:2010be,Kapustin:2013qsa,Kapustin:2013uxa,Aharony:2013hda,
DelZotto:2015isa,Sharpe:2015mja, Heckman:2017uxe, Tachikawa:2017gyf,
Cordova:2018cvg,Benini:2018reh,Hsin:2018vcg,Wan:2018bns,
Thorngren:2019iar,GarciaEtxebarria:2019caf,Eckhard:2019jgg,Wan:2019soo,Bergman:2020ifi,Morrison:2020ool,
Albertini:2020mdx,Hsin:2020nts,Bah:2020uev,DelZotto:2020esg,Hason:2020yqf,Bhardwaj:2020phs,
Apruzzi:2020zot,Cordova:2020tij,Thorngren:2020aph,DelZotto:2020sop,BenettiGenolini:2020doj,
Yu:2020twi,Bhardwaj:2020ymp,DeWolfe:2020uzb,Gukov:2020btk,Iqbal:2020lrt,Hidaka:2020izy,
Brennan:2020ehu,Komargodski:2020mxz,Closset:2020afy,Thorngren:2020yht,Closset:2020scj,
Bhardwaj:2021pfz,Nguyen:2021naa,Heidenreich:2021xpr,Apruzzi:2021phx,Apruzzi:2021vcu,
Hosseini:2021ged,Cvetic:2021sxm,Buican:2021xhs,Bhardwaj:2021zrt,Iqbal:2021rkn,Braun:2021sex,
Cvetic:2021maf,Closset:2021lhd,Thorngren:2021yso,Sharpe:2021srf,Bhardwaj:2021wif,Hidaka:2021mml,
Lee:2021obi,Lee:2021crt,Hidaka:2021kkf,Koide:2021zxj,Apruzzi:2021mlh,Kaidi:2021xfk,Choi:2021kmx,
Bah:2021brs,Gukov:2021swm,Closset:2021lwy,Yu:2021zmu,Apruzzi:2021nmk, Cvetic:2021sjm, Beratto:2021xmn,Bhardwaj:2021mzl,
Debray:2021vob, Wang:2021vki,
Cvetic:2022uuu,DelZotto:2022fnw,Cvetic:2022imb,DelZotto:2022joo,DelZotto:2022ras,Bhardwaj:2022yxj,Hayashi:2022fkw,
Kaidi:2022uux,Roumpedakis:2022aik,Choi:2022jqy,
Choi:2022zal,Arias-Tamargo:2022nlf,Cordova:2022ieu, Bhardwaj:2022dyt,
Benedetti:2022zbb, DelZotto:2022ohj, Bhardwaj:2022scy,Antinucci:2022eat,Carta:2022spy,
Apruzzi:2022dlm, Heckman:2022suy, Baume:2022cot, Choi:2022rfe,
Bhardwaj:2022lsg, Lin:2022xod, Bartsch:2022mpm, Apruzzi:2022rei,
GarciaEtxebarria:2022vzq, Cherman:2022eml, Heckman:2022muc, Lu:2022ver, Niro:2022ctq, Kaidi:2022cpf,
Mekareeya:2022spm, vanBeest:2022fss, Antinucci:2022vyk, Giaccari:2022xgs, Bashmakov:2022uek,Cordova:2022fhg,
GarciaEtxebarria:2022jky, Choi:2022fgx, Robbins:2022wlr, Bhardwaj:2022kot, Bhardwaj:2022maz, Bartsch:2022ytj, Gaiotto:2020iye,Agrawal:2015dbf, Robbins:2021ibx, Robbins:2021xce,Huang:2021zvu,
Inamura:2021szw, Cherman:2021nox,Sharpe:2022ene,Bashmakov:2022jtl, Lee:2022swr, Inamura:2022lun, Damia:2022bcd, Lin:2022dhv,Burbano:2021loy, Damia:2022rxw, Apte:2022xtu, Nawata:2023rdx, Bhardwaj:2023zix, Kaidi:2023maf, Etheredge:2023ler, Lin:2023uvm, Amariti:2023hev, Bhardwaj:2023wzd, Bartsch:2023pzl, Carta:2023bqn, Zhang:2023wlu, Cao:2023doz, Putrov:2023jqi, Debray:2023yrs, Davighi:2023luh, Acharya:2023bth, Debray:2023rlx, Debray:2023tdd} and \cite{Cordova:2022ruw} for a recent review.}

Most of these developments have centered on global symmetries, but in quantum gravity,
one expects that these symmetries are either
explicitly gauged or broken. In the Swampland program this was
recently formalized in terms of the Swampland Cobordism Conjecture, which
asserts that the bordism group of quantum gravity is trivial \cite{McNamara:2019rup}.\footnote{For recent
developments, see e.g., \cite{Debray:2021vob, Montero:2020icj, Dierigl:2020lai, Buratti:2021fiv, Blumenhagen:2021nmi,
Angius:2022aeq, Blumenhagen:2022mqw, Angius:2022mgh, Blumenhagen:2022bvh, Dierigl:2022reg, Debray:2023yrs}.}
In practice, one considers a long distance limit captured by the gravitational path
integral and then imposes specific symmetry (spacetime and internal) constraints.
Obtaining a non-trivial bordism group then amounts to the
prediction of new objects, since in the full quantum gravity there must be boundaries for the bordism classes that seemed non-trivial in the low-energy effective field theory. By now, the Cobordism Conjecture has undergone a
number of non-trivial checks in the context of supersymmetric backgrounds, and
has even been used to predict the existence of new non-supersymmetric objects
\cite{McNamara:2019rup, Dierigl:2022reg, Debray:2023yrs}.

String theory makes direct contact with both of these developments. In the
context of QFTs, string backgrounds with localized singularities in the
metric / fields / solitonic branes provide a general template for constructing
and studying a wide class of strongly coupled systems decoupled from gravity.
In this regard, it is worth noting that string theory remains \textit{the} method for
explicitly constructing interacting $D>4$ fixed
points. Indeed, the spectrum of (often supersymmetric)\ extended defects in such
systems is encapsulated in terms of the \textquotedblleft defect
group\textquotedblright\ \cite{DelZotto:2015isa, GarciaEtxebarria:2019caf, Albertini:2020mdx, Morrison:2020ool},
where branes wrapped on non-compact cycles are screened by dynamical states obtained
from branes wrapped on compact, collapsing cycles. The associated symmetry
operators which act on these defects directly encode the generalized symmetry
operators, and can be viewed as branes \textquotedblleft wrapped at
infinity\textquotedblright. Since they are infinitely far away, essentially
the only contribution they can make to the field theory is via their
topological sector (namely Wess-Zumino (WZ)\ terms). This has recently been used to exhibit
explicit examples of non-trivial fusion rules in a number of different systems
(see e.g., \cite{Apruzzi:2022rei, GarciaEtxebarria:2022vzq, Heckman:2022muc, Heckman:2022xgu}).

Given this, it is natural to ask whether the new branes predicted by the
Swampland Cobordism Conjecture also generate topological symmetry operators.
Our aim in this note will be to show that this is indeed the case for a
specific new 7-brane predicted in the context of type IIB\ dualities: the
reflection 7-brane. As found\footnote{They were also hinted at in \cite{Distler:2009ri}.} in \cite{Dierigl:2022reg, Debray:2023yrs},
these ``R7-branes'' can be viewed as a codimension-two defect of the
10D\ IIB\ supergravity. Winding once around this brane amounts to a reflection
on either the a- or b-cycle of the F-theory torus. In terms of the
IIB\ worldsheet theory, these reflections are associated with worldsheet
orientation reversal $\Omega$ and left-moving fermion parity $(-1)^{F_L}$.
This object carries a $\mathbb{Z}_{2}$ charge of the corresponding
IIB\ duality group, and as such, cannot completely \textquotedblleft
disappear\textquotedblright. Even so, there are good indications from \cite{Dierigl:2022reg}
that it is strongly coupled and potentially unstable to thickening / expansion.

That being said, wrapping such a brane \textquotedblleft at
infinity\textquotedblright\ means that it cannot contribute to the stress
energy tensor of a localized QFT\ sector. As such, we can insert these
R7-branes and deduce the corresponding symmetry operator generated by these
objects. In the context of 6D $\mathcal{N} = (2,0)$ SCFTs realized via IIB on an ADE orbifold,
we show that insertion of an R7-brane realizes a zero-form symmetry which acts as a charge conjugation
operation on the heavy string-like defects of the theory. Further
compactification to four dimensions leads to a corresponding charge
conjugation operation which can be combined with other \textquotedblleft
branes at infinity\textquotedblright\ to implement more general symmetries
such as spacetime reflections.

Beyond the case of pure geometric engineering, one can also consider D-branes
probing singularities. In some cases, the contribution from the R7-brane leads
to a large backreaction due to the putative symmetry being explicitly broken by the background, thus making it unsuitable as a topological symmetry operator. But in other cases, this can be used to engineer related charge
conjugation / reflections of the localized QFT\ sector. The basic considerations we consider here apply to other choices of non-supersymmetric branes which carry a conserved charge. In these cases, we sketch how string-realized QFTs and little string theories (LSTs) admit symmetry operators obtained from wrapping these non-supersymmetric branes ``at infinity''.

Turning the discussion around, one can argue that the existence of a suitable symmetry in the string-realized QFT implies the existence of a corresponding topological symmetry operator. This in turn \textit{requires} the existence of a suitable object which could implement this symmetry, amounting to the requirement that a suitable brane must exist. From this perspective, the main thing to verify is that such a symmetry exists in the first place. We show that for those theories with a suitable holographic dual such as the large $N$ limits of the A- and D-type 6D SCFTs with $\mathcal{N} = (2,0)$ supersymmetry, charge conjugation amounts to a reflection on $X$, the ``internal direction'' of the background $AdS_7 \times X$. One can also extend this reasoning to many other cases where one has a stringy realization of a $D$-dimensional CFT with an $AdS_{D+1}$ dual, and more broadly, it can even be applied to more general systems such as little string theories (LSTs).

\section{R7-Branes and 6D\ SCFTs}\label{sec:TWOCOMMAZERO}

We now argue that some 6D\ SCFTs have a charge conjugation symmetry which, in
the context of F-theory on an elliptically-fibered Calabi-Yau threefold, is
realized via R7-branes wrapped \textquotedblleft at infinity\textquotedblright. That being said,
we will find (by explicit analysis) that only theories with $\mathcal{N} = (2,0)$ supersymmetry
have a charge conjugation symmetry which squares to $+1$, and is implemented by the R7-brane.  This corresponds to the case of a trivial elliptic fibration.\footnote{Theories with $\mathcal{N} = (1,0)$ supersymmetry admit a charge conjugation symmetry which squares to $(-1)^{F}$,
as we explain later.}

To begin, let us recall that in F-theory on a non-compact Calabi-Yau threefold
$X\rightarrow B$, we get a 6D\ SCFT by contracting curves of the base $B$ to
zero size. D3-branes wrapped on finite volume curves provide effective strings
with tension which tends to zero as the curves' volumes vanish. In
this limit, one obtains a 6D\ SCFT. The full list of non-compact bases $B$ as
well as possible elliptic fibrations was determined in \cite{Heckman:2013pva, DelZotto:2014hpa, Heckman:2015bfa} (for reviews see \cite{Heckman:2018jxk, Argyres:2022mnu}).
The general structure of all such bases is, in the
contracting limit, given by an orbifold of the form $\mathbb{C}^{2}/\Gamma_{U(2)}$ for
$\Gamma_{U(2)}$ a finite subgroup of $U(2)$. Working in radial coordinates,
this specifies a conical geometry with an $S^{3}/\Gamma_{U(2)}$ at each radial
slice. One obtains heavy string-like defects from D3-branes wrapped on non-compact
2-cycles which extend along the radial direction and wrap a torsional 1-cycle at the boundary $S^{3}%
/\Gamma_{U(2)}$ \textquotedblleft at infinity''. Since they wrap a torsion cycle, $n$ times these defects must be trivial, which means they are charged under a $\mathbb{Z}_n$ 2-form symmetry (only discrete 2-form symmetries are possible in 6D SCFTs \cite{Cordova:2020tij}). More precisely, the spectrum of heavy string-like defects which cannot be screened by dynamical strings are classified by the
``defect group'' (see reference \cite{DelZotto:2015isa}) which is given by the
abelianization of $\Gamma_{U(2)}$, namely $H_{1}(S^{3}/\Gamma_{U(2)}%
,\mathbb{Z})= \mathrm{Ab}(\pi_{1}(S^{3}/\Gamma_{U(2)},\mathbb{Z})) = \mathrm{Ab}(\Gamma_{U(2)})$.

Thus, it should be possible to construct codimension-three topological symmetry operators that link with the above heavy string-like defects. Indeed, these can be obtained from D3-branes wrapping these same torsional 1-cycles \cite{Heckman:2022muc} in the $S^{3} / \Gamma_{U(2)}$ at infinity. Unlike the heavy string-like defects implemented by D3-branes that extend along the radial direction from infinity to the singularity where the SCFT lives, the D3-branes implementing topological operators are localized at infinity. Intuitively, this means that a small deformation cannot affect the local physics, as any backreaction must traverse an infinite distance, and their correlators can only be possibly affected by the linking with the D3-branes implementing heavy string-like defects -- precisely the definition of a topological operator.

The dualities of type IIB string theory act on these heavy string-like defects via a general
conjugation operation. As described in \cite{Tachikawa:2018njr} (see also \cite{Dabholkar:1997zd}),
the actual duality group of IIB string theory is the $\mathsf{Pin}^{+}$ double cover of $GL(2,\mathbb{Z})$. The
reflections with negative determinant given (in terms of their action on the F-theory torus)\footnote{The monodromy matrices $M$ can also be deduced from the action on the 2-form fields of type IIB that transform as a vector given by $(C_2, B_2)^T$.}
by $M_{\mathsf{F}_{L}}= \mathrm{diag}(-1,1)$ and $M_{\Omega}= \mathrm{diag}(1,-1)$, correspond
respectively to left-moving fermion parity $(-1)^{F_L}$ and worldsheet orientation
reversal $\Omega$. Each of these generators sends a D3-brane to an
anti-D3-brane:\ $\left\vert \text{D3}\right\rangle \rightarrow
\left\vert \overline{\text{D3}}\right\rangle $. This specifies a generalized charge
conjugation operation on D3-branes. In the corresponding 6D\ SCFT, this sends
each of our heavy string-like defects (obtained from wrapped D3-branes) to its
anti-string counterpart. The reflections $M_{\mathsf{F}_{L}}$ and $M_{\Omega}$ also act
non-trivially on D7-branes since we also have $\left\vert \text{D7} \right\rangle
\rightarrow\left\vert \overline{\text{D7}}\right\rangle $.

Generically, most 6D\ SCFTs do not have a charge conjugation symmetry.
Indeed, on the tensor branch it is common to encounter various 6D\ gauge theories
which are coupled to tensor multiplets. To cancel 1-loop gauge anomalies
generated by the chiral matter of the vector multiplet one must include
suitable Green-Schwarz-Sagnotti-West terms (see \cite{Green:1984bx, Sagnotti:1992qw}) which are schematically of the form $B^{a}\wedge
I^{GS}_{a}$, where $B^{a}$ is an anti-chiral 2-form field and $I^{GS}_{a}$ is a
4-form constructed via the characteristic classes of the gauge bundles. The specific
form of such couplings can be extracted from the algorithm developed in \cite{Ohmori:2014pca, Ohmori:2014kda, Heckman:2015ola},
and can also be extended to include possible couplings to
background curvatures / R-symmetries / flavor symmetries. The presence of
couplings such as $B^{a}\wedge I_{a}$ manifestly breaks the charge conjugation
symmetry since $I_{a}$ is realized via even powers of curvatures / field
strengths (and therefore, must be charge conjugation invariant), whereas $B^{a}$ is manifestly odd under the conjugation operation, since these fields couple directly to the D3-branes wrapping the non-compact 2-cycles of the ambient geometry.
In the associated F-theory background this is also expected because the gauge
theory degrees of freedom are realized via 7-branes wrapped on compact curves,
and reflections generically send 7-branes to anti-7-branes.

The exception to this general situation are those 6D\ SCFTs which have no
7-branes at all. This occurs for the celebrated $\mathcal{N} = (2,0)$ theories, as obtained
from a collection of $-2$ curves in the base with intersection form given by
the corresponding ADE\ Dynkin diagram:%
\begin{align}
A_{N}  &  :\text{ }\underset{N}{\underbrace{2,2,...,2}}\\
D_{N}  &  :\text{ }\underset{N-1}{\underbrace{2,\overset{2}{2},...,2}}\\
E_{6}  &  :\text{ }2,2,\overset{2}{2},2,2\\
E_{7}  &  :\text{ }2,2,\overset{2}{2},2,2,2\\
E_{8}  &  :\text{ }2,2,\overset{2}{2},2,2,2,2.
\end{align}

In fact, one can argue directly from the classification of superconformal algebras that only $\mathcal{N} = (2,0)$ theories could possibly have a charge conjugation symmetry represented by R7-branes. R7-branes have a worldvolume charge which is $\mathbb{Z}_2$-valued, so the charge conjugation symmetry they implement squares to $+1$. In an $\mathcal{N} = (1,0)$ theory, this is impossible, since any charge conjugation symmetry must map the supercharge $Q$ to itself. But in six Lorentzian dimensions (or Euclidean reflection-positive), the only possible charge conjugation operator that preserves chirality squares to $-1$ \cite{polchinski1998string,Ortin:2015hya}. So while there may be a charge conjugation symmetry for $\mathcal{N} = (1,0)$ theories, it is qualitatively different from the $\mathcal{N} = (2,0)$ case. In fact, this charge conjugation symmetry may be realized as simply any $\mathbb{Z}_4$ subgroup of the $SU(2)$ R-symmetry.

We now directly construct the corresponding topological symmetry operator for the $\mathcal{N} = (2,0)$ theories. This
is realized at once in terms of an R7-brane \textquotedblleft wrapped at
infinity\textquotedblright.  In terms of the local coordinates the
relevant objects are obtained as follows:%
\begin{equation}%
\begin{array}
[c]{cccccccccccc}
&  & 0 & 1 & 2 & 3 & 4 & 5 & 6 & 7 & 8 & 9\\
\text{Defect} & \text{D3} &  &  &  &  & \times & \times & \times & \times &  & \\
\text{Symm Op.} & \text{R7} & \times & \times & \times & \times & \times &  &  &
\times & \times & \times
\end{array},
\end{equation}
where the ``$0,...,5$'' directions denote the 6D\ spacetime, the ``$6$'' direction
denotes the radial direction of the base, and the ``$7,8,9$'' directions denote
the $S^{3}/\Gamma$ \textquotedblleft at infinity\textquotedblright.

Since both the $\mathsf{F}_{L}$- and $\Omega$-brane act the same way on D3-branes, we might be tempted to conclude that there is no difference in which one we use to implement this operator. However, one can wrap F1-strings or D1-branes on the non-compact 2-cycles of the ambient geometry, and this engineers pointlike defects in the $\mathcal{N} = (2,0)$ theory. The $\mathsf{F}_{L}$- and $\Omega$-branes act differently on these, mapping only F1-strings or D1-branes to their conjugates, respectively.\footnote{One can also directly see the full duality group action on objects of the theory by introducing a stack of probe D3-branes into the system. From the perspective of the 6D SCFT this is a specific real codimension-two defect which supports a supersymmetric gauge theory. In that gauge theory, the axio-dilaton descends to a marginal coupling.}  In any case, we see that much as in \cite{Heckman:2022xgu}, either R7-brane defines a real codimension-one
topological operator, and as such should be viewed as a zero-form symmetry
operator. It is in fact typical of charge conjugation that it acts non-trivially on both
pointlike and extended operators.

\subsection{Fusion Rules}\label{ssec:FUSION}

While much is still unknown about the R7-brane, general topological / anomaly
inflow arguments provide a natural candidate action for at least a subsector of the worldvolume degrees of freedom of this system \cite{Dierigl:2022reg}.
Using this, we can then consider the fusion rules for two such symmetry operators
wrapped on a 5D\ subspace of the 6D\ spacetime. For ease of exposition we focus on the $\Omega$-brane. Similar considerations apply for the
$\mathsf{F}_L$-brane.

In differential cohomology terms,\footnote{For simplicity, we take the approximation of classifying IIB charges by cohomology, but in principle one should replace this by KR-theory (see e.g., \cite{Bergman:2001rp,Distler:2009ri}) at the perturbative level and, ultimately, some unknown generalization of twisted K-theory which is covariant under S-duality. This subtlety will not affect our main conclusions.} (for physicist friendly reviews
see e.g., \cite{Apruzzi:2021nmk,Freed:2006ya,Freed:2006yc} as well as the book \cite{bar2014differential}), we can rewrite our action for the $\Omega$-brane as \cite{Dierigl:2022reg}:
\begin{equation}\label{eq:bulkR7WZ2diff}
  \int_{R7} \breve{H}_3\star \breve{f}_6+\breve{F}_5\star \breve{f}_4+\breve{H}_7\star \breve{f}_2 \,.
\end{equation}
Here, $\breve{H}_3$ and $\breve{H}_7$ denote differential characters that describe the NS 2- and 6-form fields, respectively, while $\breve{F}_5$ describes the chiral RR 4-form. The remaining differential characters $\breve{f}_k$ describe $(k-1)$-form fields that are localized on the brane worldvolume and can absorb the charges of bulk objects, such as D3-branes, ending on the R7 (see \cite{Dierigl:2022reg} for details). The product $\star$ is defined as a map
\begin{align}
\star: \quad \breve{H}^p \times \breve{H}^q \rightarrow \breve{H}^{p+q} \,,
\end{align}
producing a differential cohomology class which can naturally be integrated over $(p+q-1)$-manifolds, such as the eight-dimensional worldvolume of the R7-brane above.

Consider the 6D $\mathcal{N} = (2,0)$ SCFTs engineered from taking IIB on $B=\mathbb{C}^2/\Gamma_{SU(2)}$.
We can now expand these fields along differential cohomology classes of $S^3/\Gamma$ to obtain topological terms on the codimension-one wall, $M_5$, in the 6D spacetime. The cohomology groups of the boundary geometry are:\footnote{For ease of exposition we give the ordinary cohomology group since the lift of these generators to differential cohomology are what is relevant in the actual fusion rule calculation.}
\begin{equation}\label{eq:boundarycoho}
  H^*(S^3/\Gamma,\mathbb{Z})=\{\mathbb{Z},0,\mathrm{Ab}(\Gamma), \mathbb{Z}\} \,.
\end{equation}
Denote the generator (or generators when $\Gamma$ is of D-type) of $H^2=\mathrm{Ab}(\Gamma)$ by $t_2$ (or $t^{i=1,2}_2$ for $D_{4k}$-type) which can be lifted to a differential cohomology class $\breve{t}_2$ in the sense that it defines its characteristic class. In the notation of Section 2 of \cite{Apruzzi:2021nmk} there is a projection $I(\breve{t}_2)=t_2$. We will pay particular attention to the middle term of \eqref{eq:bulkR7WZ2diff}, returning to the other two later, and consider the following expansions (suppressing the indices in the D-type case)
\begin{align}
 \breve{F}_5= \breve{G}_3\star \breve{t}_2\\
  \breve{f}_4=\breve{g}_2 \star \breve{t}_2.
\end{align}
Reducing to $M_5$ then simply requires knowledge of the linking pairing $L_\Gamma=\int_{S^3/\Gamma}\breve{t}_2\star \breve{t}_2$ which is a $2 \times 2$ matrix in the D-type case. The resulting action on $M_5$ can now be written as
\begin{equation}\label{eq:g3cupg2}
  L_\Gamma\int_{M_5} G_3\cup g_2
\end{equation}
and if we assume that $M_5$ is torsion-free, the K\"unneth theorem implies that $G_3$ is an $\mathrm{Ab}(\Gamma)$-valued 3-form which is hardly surprising since this is precisely the background field for the 2-form symmetry of the 6D $\mathcal{N} = (2,0)$ theory. The path integral of this 5D TFT can be written as
\begin{equation}\label{eq:pathinttft5}
 \mathcal{P}_{2}(M_5) \equiv \int D g_2 \, e^{2\pi i L_\Gamma\int_{M_5} G_3\cup g_2} = \sum_{\Sigma_3\in H_3(M_5,\mathrm{Ab}(\Gamma)) }e^{2\pi i L_\Gamma \int_{\Sigma_3} G_3}.
\end{equation}
where again we point out that we have suppressed the extra indices in $L^{ij}_\Gamma$ for the $D_{4k}$ case. Since $e^{2\pi i L_\Gamma \int_{\Sigma_3} G_3}$ can be interpreted as a symmetry operator for $\mathrm{Ab}(\Gamma)^{(2)}$, we see that we are gauging this symmetry along $M_5$. In the language of \cite{Roumpedakis:2022aik}, this is a 1-gauging of a 2-form symmetry. Returning to the other two terms in \eqref{eq:bulkR7WZ2diff}, we see that those produce 1-gaugings of $\mathrm{Ab}(\Gamma)^{(4)}$ and $\mathrm{Ab}(\Gamma)^{(0)}$ symmetries, denoted as $\mathcal{P}_4$ and $\mathcal{P}_0$, respectively, whose charged operators arise from wrapping NS5-branes and F1-strings on relative 2-cycles which, topologically, are cones over the boundary 1-cycles. We then can write our charge conjugation operator as
\begin{equation}\label{eq:chargeconjtotal}
  \mathcal{U}_{\Omega}(M_5)=\mathsf{C}\cdot \mathcal{P}_{0}\cdot \mathcal{P}_{2}\cdot  \mathcal{P}_{4}.
\end{equation}
where $\mathsf{C}$ is the more elementary charge conjugation which simply acts on the operators of the 6D $\mathcal{N} = (2,0)$ theory in the form we mentioned above. We have that $\mathsf{C}^2=1$ because the R7 monodromy matrix, as an element in $GL(2,\mathbb{Z})$ lifts to an order-two element in $GL^+(2,\mathbb{Z})$ \cite{Debray:2021vob}. As discussed in \cite{Heckman:2022muc}, the operators $\mathcal{P}_{k}$ which enact a $p$-gauging of a $k$-form symmetry satisfy $\mathcal{P}^2_k=\mathcal{P}_k$, i.e., they are projection operators onto sectors where the flux being gauged vanishes. This does not have a well-defined inverse which is the sense in which our charge conjugation operator engineered from the R7-brane, $\mathcal{U}_{\Omega}$, is non-invertible. So in summary, the fusion rules of $\mathcal{U}_{\Omega}$ with itself are summarized as
\begin{align}\label{eq:r7fusion}
  &\mathcal{U}^2_{\Omega}=\mathcal{U}^\dagger_{\Omega}\cdot\mathcal{U}_{\Omega}=\mathcal{P}_0\cdot \mathcal{P}_2\cdot \mathcal{P}_4.
\end{align}

\begin{figure}[t!]
\begin{center}
\includegraphics[scale = 0.9]{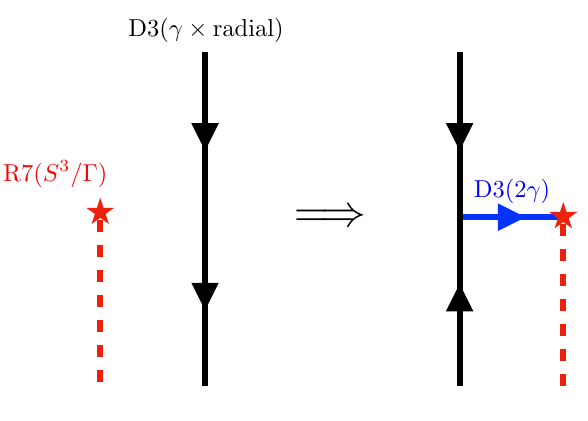}
\caption{Here we illustrate the effect of dragging a D3-brane (oriented black line) through an R7-brane (red star) whose cut associated to the monodromy action $C_4\rightarrow -C_4$ is denoted by the dashed red line. We also denote the submanifolds of $\mathbb{C}^2/\Gamma$ wrapped by these branes, here $\gamma \in H_1(S^3/\Gamma)$ is the torsion 1-cycle, relevant to constructing the 0-form charge conjugation operator for 6D SCFTs.}
\label{fig:hweffect}
\end{center}
\end{figure}

\begin{figure}[h!]
\begin{center}
\includegraphics[scale = 0.9]{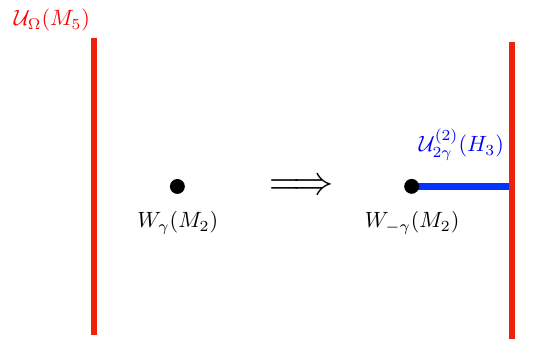}
\caption{Spacetime view of the Hanany-Witten process illustrated in Figure \ref{fig:hweffect} where we now indicate the spacetime submanifolds where these operators are supported. As the charged string defect operator $W_\gamma(M_2)$ passes through the charge conjugation operator $\mathcal{U}_{\Omega}(M_5)$, the 2-form symmetry operator $\mathcal{U}^{(2)}_{2\gamma}$ is created and stretches between $W_{-\gamma}(M_2)$ and $\mathcal{U}_{\Omega}(M_5)$.}
\label{fig:hweffectspacetime1}
\end{center}
\end{figure}

We now consider the effect of passing a string defect operator $W_\gamma(M_2)$ with charge\footnote{Technically speaking, we should write $\widetilde{\gamma}\in (\mathrm{Ab}(\Gamma)^{(2)})^\vee$ where $^\vee$ denotes Pontryagin dual and $\widetilde{\gamma}$ pairs perfectly with $\gamma$, but we choose not to overload the notation.} $\gamma\in \mathrm{Ab}(\Gamma)^{(2)}$ through $\mathcal{U}_{\Omega}(M_5)$. This can be determined by passing a D3-brane through an R7 as in Figure \ref{fig:hweffect}. We see that two D3-branes (with the orientations illustrated) emanate from the R7 as required for consistency with charge conservation. This Hanany-Witten effect is similar to the usual case of passing $[p,q]$ strings / 5-branes through supersymmetric 7-branes.\footnote{The $[p,q]$ strings / 5-branes also experience a Hanany-Witten effect for R7-branes, which, for example is non-trivial for $p\neq 0$ for the $\Omega$-brane. The relevance of Hanany-Witten moves in the study of symmetry operators was noted in \cite{Apruzzi:2022rei} and was further explored in \cite{Heckman:2022muc}.} We see also from Figure \ref{fig:hweffect} that if we regard the vertical direction as the radial direction of $\mathbb{C}^2/\Gamma$ with $r=0$ indicating the bottom of the figure, then the ending D3-brane created from the Hanany-Witten-like move is located at the asymptotic boundary. This D3-brane is nothing other than the symmetry operator associated to $\mathrm{Ab}(\Gamma)^{(2)}$, which we denote by $\mathcal{U}^{(2)}_{2\gamma}$. The worldvolume of this D3 is $H_3\times \{ 2\gamma \}$ where $H_3$ is a 3-manifold in the 6D spacetime such that $\partial H_3=M_2\coprod \overline{M_2}$ see Figure \ref{fig:hweffectspacetime1}, and we use $2\gamma$ to denote a 1-cycle in $S^3/\Gamma$ with such a charge in $H_1 (S^3 / \Gamma )$. We thus have the fusion rule
\begin{equation}\label{eq:hweffect}
  \mathcal{U}_{\Omega}\cdot W_\gamma(M_2)=W_{-\gamma}(M_2)\cdot \mathcal{U}^{(2)}_{2\gamma}(H_3)\cdot\mathcal{U}_{\Omega}.
\end{equation}
This effect of a creation of another topological symmetry operator when passing a heavy operator through a 0-form symmetry operator is a common feature of non-invertible symmetries. This notably happens when passing (dis)order operators through the Kramers-Wannier duality defect in the Ising model \cite{Frohlich:2004ef, Frohlich:2006ch, Frohlich:2009gb, Chang:2018iay} (see also \cite{Verlinde:1988sn, Aasen:2016dop, Freed:2018cec}). Note that for $D_{4k}$ type theories, the action on $W_{\gamma}(M_2)$ is trivial since the charge of $\gamma$ is labeled by $\mathbb{Z}_2\times\mathbb{Z}_2$.

The fusion rule \eqref{eq:hweffect} can simplify after one chooses a polarization for the 6D SCFT defect group, or equivalently, gauge a maximal non-anomalous subgroup of $\mathrm{Ab}(\Gamma)^{(2)}$ such that a given $W_\gamma(M_2)$ is a genuine defect operator while some $\mathcal{U}^{(2)}_{2\gamma'}$  is summed over the entire spacetime. In this case, $\mathcal{U}^{(2)}_{2\gamma'}(H_3)$ would no longer appear in the fusion rule since it is projected out of the theory.\footnote{Note that this is not always be possible as for instance when $|\mathrm{Ab}(\Gamma)^{(2)}|$ is a square-free integer.} For an illustrative example, take the type $A_{p^2-1}$ 6D $(2,0)$ theory where $\mathrm{Ab}(\Gamma)^{(2)}=\mathbb{Z}^{(2)}_{p^2}$. A priori this is a relative theory and we can form an absolute 6D SCFT by gauging $\mathbb{Z}^{(2)}_p\subset \mathbb{Z}^{(2)}_{p^2}$. If we denote $\gamma$ as a generator of $\mathbb{Z}^{(2)}_{p^2}$, the gauging implies that we sum over networks of topological operators $\mathcal{U}^{(2)}_{p\gamma}$ such that $p\gamma\in \mathbb{Z}^{(2)}_p\subset \mathbb{Z}^{(2)}_{p^2}$. The gauged theory has the topological operators $\mathcal{U}^{(2)}_{\gamma\; \mathrm{mod}\; p}$ that generate the remaining $\mathbb{Z}^{(2)}_p$ symmetry (we leave the $\mathrm{mod}\; p$ implicit in what follows). From the string defect perspective, we start in the relative 6D theory with non-genuine defects $W_{\widetilde{\gamma}}(M_2)\cdot\mathcal{U}_{\gamma}(M_3)$ where $\widetilde{\gamma}$ generates the Pontryagin dual group $(\mathbb{Z}^{(2)}_{p^2})^\vee$, $\widetilde{\gamma}(\gamma)=1/p^2 \; \mathrm{mod}\; 1$, and $\partial M_3=M_2$. After gauging $\mathbb{Z}^{(2)}_p$, we have genuine defects $W_{p\widetilde{\gamma}}(M_2)\in (\mathbb{Z}^{(2)}_p)^\vee\subset (\mathbb{Z}^{(2)}_{p^2})^\vee$, while all other defects (i.e. ones non-trivial in $(\mathbb{Z}^{(2)}_{p^2})^\vee/(\mathbb{Z}^{(2)}_p)^\vee$) are non-genuine. We now observe what happens when we drag a genuine and non-genuine defect across a charge conjugation operator $\mathcal{U}_{\Omega}$. We see that the genuine defect no longer has a topological operator attached because equation \eqref{eq:hweffect} now reads
\begin{equation}
  \mathcal{U}_{\Omega}\cdot W_{p\widetilde{\gamma}}(M_2)=W_{-p\widetilde{\gamma}}(M_2)\cdot \mathcal{U}_{2p\gamma}(H_3)\cdot\mathcal{U}_{\Omega}
\end{equation}
but $\mathcal{U}_{2p\gamma}=1$ in the gauged theory so there is no extra topological operator attached. Meanwhile, the non-genuine defect has its attached topological operator altered by $\mathcal{U}_{\gamma}\mapsto \mathcal{U}_{-\gamma}$. In other words, the right-hand side of the fusion rule would automatically be accompanied by an extra $\mathcal{U}_{2\gamma}(H_3)$.

From this example, we then see that appearance of condensation operators in the definition of $U_\Omega(M_5)$ in \eqref{eq:chargeconjtotal} also follows from bottom-up considerations. This is because we are allowed to spontaneously create open topological defects of the form $U_{2\gamma}(N_3)$ on its worldvolume where $\partial N_3\subset M_5$. This follows from moving $W_{\widetilde{\gamma}}$ across $U_\Omega(M_5)$ and back again which means that a network of $U_{2\gamma}(N_3)$ is implicitly summed on the charge conjugation worldvolume $M_5$. For a similar point, see Figure 5 of \cite{Choi:2021kmx} which shows this creation property for duality defects.

For completeness, we also mention the analogous fusion rules relevant for the action of the R7 charge conjugation operator on the local operators and 4-manifold defects charged under $\mathrm{Ab}(\Gamma)^{(4)}$ and $\mathrm{Ab}(\Gamma)^{(0)}$ in the obvious notational adaptations
\begin{align}\label{eq:hweffect2}
  & \mathcal{U}_{\Omega}\cdot W_\gamma(x)=W_{-\gamma}(x)\cdot \mathcal{U}^{(0)}_{2\gamma}(H_1)\cdot \mathcal{U}_{\Omega} \\
  & \mathcal{U}_{\Omega}\cdot W_\gamma(M_4)=W_{-\gamma}(M_4)\cdot \mathcal{U}^{(4)}_{2\gamma}(H_5)\cdot \mathcal{U}_{\Omega}.
\end{align}
Similar remarks related to the simplification after choosing the polarization apply to these symmetries as well.

\subsection{Using Holographic CFTs to Predict Cobordism Defects}\label{ssec:HOLO}

Up to this point, we have \textit{assumed} the existence of the R7-brane and have shown that it admits a natural interpretation as a
charge conjugation symmetry operator in certain 6D SCFTs. We now turn the discussion around and use holography to argue for the \textit{existence} of this cobordism defect.

Along these lines, the main idea will be to first show that for 6D $\mathcal{N} = (2,0)$ SCFTs with a semi-classical holographic dual,
the gravity dual admits a discrete symmetry which we shall interpret as a charge conjugation symmetry in the 6D SCFT. As such, there must exist a corresponding codimension one topological symmetry operator. Proceeding back from the M-theory realization to the F-theory realization, this amounts to a complementary expectation that there must exist a corresponding object in IIB which implements this symmetry operator: this is nothing but the R7-brane.

To proceed, recall that there are well-known holographic duals for some of the 6D\ SCFTs just
considered. For example, for the A-type $\mathcal{N} = (2,0)$ theories, we can start from $N$ coincident M5-branes in flat
space, we reach the gravity dual given by M-theory on $AdS_{7}\times S^{4}$
with $N$ units of 4-form flux through the $S^{4}$ (see e.g., \cite{Maldacena:1997re}).
Similar considerations hold for the D-type theories, where the holographic dual is $AdS_7 \times \mathbb{RP}^4$.

All the states, operators and symmetries that we found above, including the charge conjugation symmetry, must be apparent in the holographic dual. In this picture, the string defects obtained from D3-branes wrapping non-compact 2-cycles are represented by M2-branes attached to the boundary of the holographic dual. The charge conjugation symmetry is implemented in terms of the Pin$^+$ symmetry of M-theory \cite{Witten:1995em,Witten:1996md,Dabholkar:1997zd,Witten:2016cio,Tachikawa:2018njr}, under which the M-theory 3-form $C_3$ transforms as a pseudo 3-form. What this means is that, in a compactification of the form $AdS_7\times X_4$, a reflection of an $AdS_7$ coordinate is not a symmetry of the theory, because the $G_4$ flux threading $X_4$ flips sign (and thus changes the vacuum), but a reflection on $X_4$ (if there is such a symmetry available) will flip both $G_4$ and the sign of the volume form, being a symmetry of the theory. Indeed, there are M-theory backgrounds which are holographically dual to $\mathcal{N} = (2,0)$ theories in the large $N$ limit of the A- and D-type theories. These involve an $X_4$ which is either $S^4$ or $\mathbb{RP}^4$, and both preserve discrete symmetries which in the 6D SCFT specify a charge conjugation which squares to $+1$ in the 6D SCFT.\footnote{To be even more concrete, let us illustrate how some examples of such reflections are implemented on $S^4$ and $\mathbb{RP}^{4}$. Starting with an $S^{4}$ of radius $L$, we view it as the real hypersurface $(x_1)^2 + (x_2)^2 + (x_3)^2 + (x_4)^2 + (x_5)^2 = L^2$ in $\mathbb{R}^{5}$. The reflection $(x_1,x_2,x_3,x_4,x_5) \mapsto (-x_1,x_2,x_3,x_4,x_5)$ induces a corresponding reflection on the $S^4$. Other reflections are obtained by performing a rotation on the $S^4$. We reach $\mathbb{RP}^{4}$ by quotienting $S^{4}$ by the anti-podal map $(x_1,x_2,x_3,x_4,x_5) \mapsto (-x_1,-x_2,-x_3,-x_4,-x_5)$. This still retains a $\mathbb{Z}_2$ symmetry given by reflection of one of the ambient $\mathbb{R}^5$ coordinates, so this descends to a charge conjugation symmetry of the D-type theory.}

In fact, at this point, one may very well flip the logic. Using the fact that the $\mathcal{N} = (2,0)$ theory has a charge conjugation symmetry that squares to $+1$, we \textit{predict} the existence of the R7-brane as the object that realizes the corresponding topological operator in the IIB description. The R7-brane was originally described in \cite{Dierigl:2022reg, Debray:2023yrs} as a consequence of the Cobordism Conjecture -- but from this perspective, its existence is required by holography and the standard IIB description of the $\mathcal{N} = (2,0)$ theory. In short, one can use concrete holographic constructions to provide evidence for some of the non-supersymmetric objects predicted by the Cobordism Conjecture!

The considerations just presented also apply to many other situations, including beyond the AdS/CFT correspondence. For example, the holographic dual of a little string theory is (when it exists), flat space with a linear dilaton profile \cite{Aharony:1998ub}. In such situations one can consider discrete reflection-type symmetries of the ``internal'' directions. This also applies to the near horizon limits of various black (and grey) objects in gravity. In short, the existence of a discrete symmetry in a holographic (but not necessarily $AdS$) dual provides evidence for a corresponding topological symmetry operator which must be implemented by a suitable object.

\section{Compactification and Further Reflections}\label{sec:REFLECTIONS}

Starting from the 6D\ $\mathcal{N}=(2,0)$ theories, one reaches a range of
4D\ SCFTs with $\mathcal{N}\geq1$ supersymmetry by compactifying further on a
genus $g$ Riemann surface with punctures (see e.g., \cite{Witten:1997sc, Gaiotto:2009we, Bah:2012dg}).

It is natural to ask whether the R7-brane still implements a charge
conjugation topological operator in this compactified theory. Although at first it would seem that the answer is always affirmative, since one can just wrap the 6D topological defect on the Riemann surface, additional ingredients such as a non-trivial
flavor or R-symmetry bundle can still end up breaking the charge conjugation symmetry of the parent 6D theory. In such situations, one might still have a charge conjugation symmetry but it will have to be combined with additional discrete symmetry actions.

One should expect to have a charge conjugation symmetry in many cases. For example,
this is the case for 4D $\mathcal{N}=2$ supersymmetric theories. The question is whether the charge conjugation symmetry thus obtained in four dimensions can be directly traced back to the 6D $\mathsf{C}$ that squares to $+1$ and that we described above. When the reduction is on $T^2$, to produce an $\mathcal{N}=4$ theory, this is automatically the case, and more generally, any toroidal compactification of the 6D $\mathcal{N} = (2,0)$ SCFT will inherit a charge conjugation symmetry. However, when the compactification is on another genus $g \neq 1$ Riemann surface, the non-trivial R-symmetry bundle used to implement a partial topological twist of the theory will generically break the charge conjugation symmetry, and the same will happen when punctures are included. Moreover, in the case of 4D\ $\mathcal{N}=1$ theories, the presence of background
curvatures / flavor fluxes will generically lead to a chiral spectrum and
broken charge conjugation symmetry (for example, a 6D\ hypermultiplet in the
presence of a background flavor flux will descend to a 4D Weyl fermion).

We now briefly comment on spacetime reflection symmetries. Unlike ordinary symmetries, spacetime symmetries (and in particular, reflections) are not captured by simple topological operators. The only meaning of a reflection in a QFT is that the QFT makes sense on non-orientable manifolds (see \cite{McNamara:2022lrw} for a recent discussion of this point). Non-orientability is detected by the first Stiefel-Whitney class $w_1$; if we transport any operator along a closed path in the $\mathbb{Z}_2$ cycle dual to $w_1$, it will come back ``reflected'' to the starting point. One can take the point of view that this is because in going around the cycle it ``crossed'' a topological defect inducing a reflection (see \cite{McNamara:2022lrw} for a detailed exposition of this point), but such notions can be misleading since one cannot ``insert'' the operator in any orientable manifold. In cases where both charge conjugation and reflection symmetries are present, one may construct, in the restricted sense described above, a time-reversal operator. This provides a top down route to implementing various time-reversal symmetry defects of the sort considered in \cite{Choi:2022rfe}.

\subsection{Other Brane Systems}\label{ssec:OTHERBRANES}

So far, our discussion has primarily focused on the case of SQFTs engineered
purely from singular background geometries. One can also consider D-brane
probes of a singularity, and ask whether the R7-brane introduces a charge
conjugation operation in this setting as well. In some cases, we find that the R7-brane does not implement
a charge conjugation symmetry operator, and so we instead seek an alternative,
which we explicitly provide in various cases.

\begin{figure}[t!]
\begin{center}
\includegraphics[scale = 0.8]{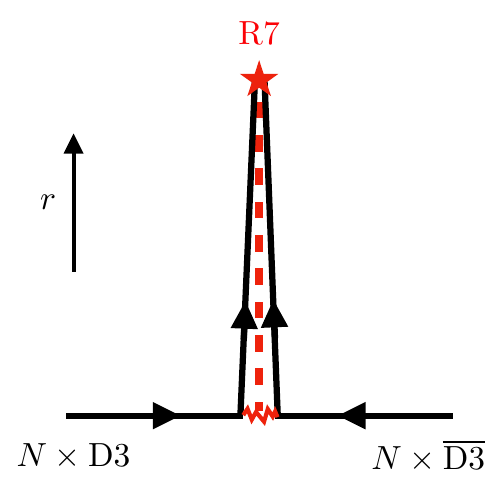}
\caption{Depiction of $N$ D3-branes (at $r=0$) in the presence of an R7-brane (at
$r=\infty$). Because the R7-brane sends D3-branes to anti-D3-branes, there is a large
jump in the flux and a number of D3-branes extend out from the D3 to the
R7-brane at infinity. The jump in the flux emanates from the branch cut
(dashed red). In this case the R7-brane does not produce a topological
operator due to the significant change to the QFT\ sector. Rather, it becomes a non-supersymmetric interface between $\mathcal{N}=4$ SYM to itself.}
\label{fig:R7D3}
\end{center}
\end{figure}

It is instructive to observe that not all R7-branes can be introduced as
topological operators in such constructions. For example, precisely because
the $\mathsf{F}_{L}$-brane acts via $\left\vert \text{D}p\right\rangle \rightarrow\left\vert
\overline{\text{D}p}\right\rangle $, this leads to a rather dramatic jump in the
asymptotic profile of the corresponding RR\ flux at the boundary of the
background spacetime. Placing the $\mathsf{F}_{L}$-brane at infinity then leads to a
large backreaction in which the RR\ flux jumps from $N$ to $-N$. See Figure
\ref{fig:R7D3} for a depiction in the case of D3-branes.

The $\Omega$-brane introduces no such issues for D1- and D5-branes, but again
sends $\left\vert \text{D3}\right\rangle \rightarrow\left\vert \overline
{\text{D3}}\right\rangle $ and $\left\vert \text{D7}\right\rangle \rightarrow\left\vert
\overline{\text{D7}}\right\rangle $. As such, we conclude that a charge conjugation
operator may be realized in the D1- and D5-brane gauge theories via $\Omega
$-branes, but not in these other systems. Lastly, one
can also consider the S-dual brane configurations, and in such situations the
roles of the $\mathsf{F}_{L}$- and $\Omega$-brane are reversed.

As an illustrative example, consider type IIB on $\mathbb{R}^{5,1}%
\times\mathbb{C}^{2}$ with $N$ D5-branes filling the first factor. In this
system, we have Wilson line defects as obtained from F1-strings which run
along the radial direction of $\mathbb{C}^{2}= \mathrm{Cone}(S^{3})$, and 't Hooft
\textquotedblleft membranes\textquotedblright, from D3-branes which wrap the same radial
direction and fill a three-dimensional subspace of $\mathbb{R}^{5,1}$. The
topological operator which implements charge conjugation is given by an
$\Omega$-brane wrapped on the boundary $S^{3}=\partial\mathbb{C}^{2}$. Indeed,
observe that both the F1-string and D3-brane are conjugated to their
anti-brane counterparts upon passing through the corresponding topological
defect. Wrapping on a $T^{2}$ and T-dualizing, we get a 4D gauge theory on the
worldvolume of a D3-brane. In this setting, the wrapped D3-brane descends to a
D1-brane, namely the 't Hooft line defect of the 4D\ theory. Observe also that
T-duality must act non-trivially on the wrapped R7-brane to realize charge
conjugation in this new theory.

\paragraph{D3-Brane Stack}
Recently it was shown that for D3-brane probes of geometry, wrapping 7-branes with a constant axio-dilaton profile ``at infinity'' provides a natural way to implement and unify various approaches to the duality defects of \cite{Kaidi:2021xfk, Choi:2021kmx} from a top down vantagepoint \cite{Heckman:2022xgu}. A natural candidate for a charge conjugation operator for a stack of $N$ D3-branes realizing an $\mathcal{N}=4$ $\mathfrak{su}(N)$ gauge theory is the $I^*_0$ 7-brane\footnote{In perturbative string language, this is a collection of 4 D7-branes coincident with an O7$^-$ plane.} wrapped along the boundary $S^5$ and a codimension-one manifold in the D3 worldvolume. An important feature of the $I^*_0$ 7-brane compared with other constant axio-dilaton 7-branes is that it does not fix a specific value of the axio-dilaton.\footnote{The Weierstrass model for an $I_{0}^{\ast}$ singularity is $y^2 = x^3 + f_0 z^2 x + g_0 z^3$. Tuning $f_0$ and $g_0$, one can reach any desired value of the axio-dilaton.}

The monodromy matrix for this 7-brane is given by
\begin{equation}\label{eq:i0starmono}
 C\equiv  \begin{pmatrix}
    -1 & 0 \\
    0 & -1
  \end{pmatrix}\in GL(2,\mathbb{Z})
\end{equation}
which in particular sends F1- and D1-strings to their anti-string counterparts. Since an F1- / D1-string stretching from infinity and ending on the D3 stack is a fundamental Wilson / 't Hooft line we see that \eqref{eq:i0starmono} indeed specifies a charge conjugation. The directions of the various branes in this scenario are as follows:
\begin{equation}%
\begin{array}
[c]{cccccccccccc}
&  & 0 & 1 & 2 & 3 & 4 & 5 & 6 & 7 & 8 & 9\\
\text{QFT Worldvolume} & \text{D3} & \times & \times & \times & \times &  &  &  &  & &            \\
\text{Defect} & \text{F1 or D1} &  & \times &  &  & \times &  &  &  &  & \\
\text{Symm Op.} & I^*_0 \; \text{7-brane} & \times & \times & \times &  &  & \times & \times & \times & \times & \times
\end{array}
,
\end{equation}
where the directions ``0,...,3'' represent the D3 worldvolume, the ``4'' direction is the radial direction of the transverse $\mathbb{C}^3$, and ``5,...9,'' represent the asymptotic $S^5$ boundary. Similar to Section \ref{ssec:FUSION}, we denote the $M_3\times S^5$ as the total worldvolume of the 7-brane which produces a topological charge conjugation symmetry operator $\mathcal{U}_{I^*_0}(M_3)$. A key feature that differentiates this charge conjugation operator from those engineered from R7-branes is that $C$ in \eqref{eq:i0starmono} lifts to an order-four element $\widehat{C}$ in $GL^+(2,\mathbb{Z})$ which satisfies $\widehat{C}^2=(-1)^F$, whereas the lift of the R7 monodromy will be an order-two element which squares to the identity due to the $\mathsf{Pin}^+$ condition. For an explicit presentation of generators and relations of $GL^+(2,\mathbb{Z})$ see \cite{Debray:2021vob}.

Since $\mathcal{N}=4$ SYM can be obtained from dimensional reduction of the 6D $(2,0)$ $A_{N-1}$ theory on $T^2$ \cite{Vafa:1997mh}, we  expect to have a charge conjugation operator which squares to $+1$. To construct it in the D3-brane system, one may combine the charge conjugation action $\mathsf{C}$ defined above with any order-four element of the $SU(4)$ R-symmetry group. The resulting operator, which we will call $\widehat{\mathsf{C}}$, will act on F1- and D1-strings as above, while not commuting with the R-symmetry; these are precisely the properties of the 6D charge conjugation operator that we discussed previously.

To summarize then, the $I^*_0$ on $M_3\times S^5$ engineers the operator
\begin{equation}\label{eq:}
  \mathcal{U}_{I^*_0}(M_3)=\widehat{\mathsf{C}}\cdot \mathrm{TFT}_3
\end{equation}
where $\widehat{\mathsf{C}}^2=(-1)^F$ and $\mathrm{TFT}_3$ is a 3D TFT living on the worldvolume of the topological operator. From the WZ term on the $I^*_0$ worldvolume,
\begin{equation}\label{eq:so8stackworldvolume}
  S_{WZ,I^*_0}\supset \int_{M_3\times S^5}C_4 \,\mathrm{Tr}F_{\mathfrak{so}(8)}^2,
\end{equation}
we find that the TFT is simply a level $N$ Chern-Simons theory with gauge algebra $\mathfrak{so}(8)$. As in the case of $\mathcal{U}_{\Omega}$, we similarly obtain a Hanany-Witten effect whereby a topological surface operator attaches to a line operator after dragging it through $\mathcal{U}_{I^*_0}(M_3)$.

Finally, note that clearly these remarks generalize straightforwardly to constructing charge conjugation operators of SCFTs engineered from D3-brane probes of a Calabi-Yau 2-fold singularity. The non-trivial boundary topology can generally cause the fusion rules to become far richer as the bevy of terms in the Wess-Zumino action of the $I^*_0$ 7-brane other than \eqref{eq:so8stackworldvolume} will also have non-trivial KK-reductions.\footnote{See for instance Section 5 of \cite{Heckman:2022xgu} which studied the dimensional reduction of various IIB 7-branes on $S^5/\Gamma$ in order to calculate the fusion of duality defects for 4D $\mathcal{N}=1$ SCFTs engineered from D3-branes probing $\mathbb{C}^3/\Gamma$. From that point of view, charge conjugation can be seen as a special case of a duality defect.}

\subsection{Symmetry Operators from Other Non-supersymmetric Branes}\label{ssec:OTHERSYMOPS}

We now comment on how various non-supersymmetric branes in heterotic and Type I string theories can be used to construct topological symmetry operators for various field theories and Little String Theories (LSTs). The Type I non-supersymmetric branes were first discussed long ago (see e.g. \cite{Sen:1999mg, Schwarz:1999vu} for reviews and \cite{Kaidi:2019pzj,Kaidi:2019tyf,Witten:2023snr} for recent discussions of these branes from a worldsheet point-of-view) and admit a KO-theory classification which is roughly equivalent to the topological configurations of the gauge field associated to non-trivial homotopy groups $\pi_*(SO(32))$ \cite{Witten:1998cd}. Meanwhile, non-supersymmetric branes in heterotic string theories were recently discovered\footnote{The authors of \cite{Kaidi:2023tqo} point out that the non-BPS 0-brane they discuss is an endpoint for the $Spin(32)/\mathbb{Z}_2$ heterotic string, as initially proposed in \cite{Polchinski:2005bg}.} in \cite{Kaidi:2023tqo,Kaidi:2023tqoappear}. While our presentation is not exhaustive, our aim is to highlight some of the minimal settings in which these branes play a role as symmetry operators. These will be broadly applicable to geometric and brane engineering of QFTs or LSTs since these branes do not act on RR $p$-form potentials nor on the NSNS 2-form, and thus will not cause a large backreaction as we saw in Figure \ref{fig:R7D3}. As for the non-supersymmetric branes not mentioned in this subsection, which include the heterotic 4-brane and Type I D8-brane, we leave the exploration of their utility as symmetry operators for future work. Again, this Section can also be read ``backwards'', in the sense that the fact that the symmetry operators must exist in the corresponding worldvolume theories provides indirect evidence for the existence of the corresponding non-supersymmetric branes in the dual quantum theory of gravity.

\paragraph{$\left(E_8\times E_8\right) \rtimes \mathbb{Z}_2$ Heterotic 7-brane}
The heterotic 7-brane introduced in \cite{Kaidi:2023tqo} is characterized by having a monodromy that exchanges the two $E_8$ factors of the gauge group. In other words, there is a non-trivial Wilson line for the $\mathbb{Z}_2$ outer automorphism factor in $\left(E_8\times E_8\right) \rtimes \mathbb{Z}_2$. Wrapping this 7-brane along the asymptotic spatial directions would then be a 0-form symmetry that exchanges the two $E_8$ factors in a flavor group associated to some localized degrees of freedom.

A natural candidate for a physical system that may realize this 7-brane as a symmetry operator are small heterotic instantons arranged such that the instanton numbers are the same for both $E_8$ factors. In heterotic M-theory language,\footnote{In this duality frame, the non-supersymmetric 7-brane uplifts to pure geometry and is associated with reflection along the interval direction.} this amounts to considering the same number $N$ of parallel M5-branes arranged symmetrically between the two $E_8$ walls. As in \cite{Intriligator:1997dh, Bhardwaj:2015oru}, we can consider a gravitational decouping limit to isolate these 6D degrees of freedom such that the size of the interval between the two $E_8$ walls remains fixed but is much larger than the ten-dimensional Planck length,  which engineers a 6D Little String Theory (LST). If we consider $N$ M5-branes, this engineers a rank-$N$ E-string LST whose tensor branch is captured in the dual F-theory geometry as follows (where the number denote the self-intersection numbers of 2-cycles in the dual F-theory geometry):
\begin{align}
\textnormal{Rank-$N$ E-string LST}  &  :\text{ }[E_8]\; 1,\underset{N-2}{\underbrace{2,2,...,2}} ,\; 1\; [E_8]
\end{align}
Of the $N$ compact curves, it is only possible to blow down $N-1$ of them with the volume of the remaining curve corresponding with the intrinsic length scale of the LST. The non-supersymmetric 7-brane then engineers a 0-form symmetry exchanging the $E_8$ flavor factors only for a subregion in the LST tensor branch that respects this symmetry. For example, if we take $N$ to be even and are at a tensor branch location such that $N/2$ M5-branes are at one $E_8$ wall and $N/2$ at the other, then the 7-brane is indeed a symmetry operator. Under the RG flow to the IR we have
\begin{equation*}
\textnormal{(Rank-$N$ E-string LST)}\; \rightarrow\; \textnormal{(Rank-$N/2$ E-string SCFT) $\oplus$ (Rank-$N/2$ E-string SCFT)}
\end{equation*}
where the right-hand side is a direct sum of two identical Rank-$N/2$ E-string SCFTs and the 0-form symmetry in the IR simply exchanges these two factors.

Similar remarks equally hold if we take four of the spatial directions of the $\left(E_8\times E_8\right) \rtimes \mathbb{Z}_2$ heterotic string theory to be an ADE singularity $\mathbb{C}^2/\Gamma_{ADE}$ and consider small instanton probes thereof \cite{Seiberg:1996vs,Ganor:1996mu,Witten:1995gx}. These are known as orbi-instanton LSTs, and on a partial tensor branch are characterized by the F-theory geometry
\begin{align}
\textnormal{Orbi-instanton Rank-$N$ E-string LST}  &  :\text{ }[E_8]\; \overset{\mathfrak{g}_{ADE}}{1},\underset{N-2}{\underbrace{\overset{\mathfrak{g}_{ADE}}{2},\overset{\mathfrak{g}_{ADE}}{2},...,\overset{\mathfrak{g}_{ADE}}{2}}},\; \overset{\mathfrak{g}_{ADE}}{1}\; [E_8]
\end{align}
where the notation $\overset{\mathfrak{g}_{ADE}}{n}$ denotes a $(-n)$-curve with a 7-brane hosting gauge degrees of freedom with Lie algebra $\mathfrak{g}_{ADE}$ wrapping it.

Finally, we mention that this 7-brane would engineer a symmetry operator on a 2D $\mathcal{N} = (0,1)$ SCFT associated to the heterotic string itself. This 0-form symmetry of course acts as an outer automorphism on the momentum lattice associated the internal left-moving $T^{16}$ geometry.

\paragraph{7-Brane of Type I String and 6-Brane of Heterotic String}
Another set of non-supersymmetric branes that can easily be interpreted in terms of symmetry operators are the $\mathbb{Z}_2$-valued 7-brane in Type I string theory and the $\mathbb{Z}_2$-valued 6-brane in heterotic $Spin(32)/\mathbb{Z}_2$ string theory. The former is associated with a $Spin(32)/\mathbb{Z}_2$ gauge bundle such that we have a non-trivial Wilson line along the transverse angular $S^1$ direction. In particular, the non-supersymmetric 0-brane of Type I (which is S-dual to the massive spinor state in perturbative heterotic string theory) is a $Spin(32)/\mathbb{Z}_2$ spinor state which has a non-trivial monodromy around this 7-brane \cite{Witten:1998cd,Gukov:1999yn}. In other words, the 7-brane is characterized by a Wilson line in the center of $Spin(32)/\mathbb{Z}_2$ and winding around the bounding $S^1$ transverse to the 7-brane.  As for the heterotic 6-brane, this is characterized by a non-trivial integral of the second Stiefel-Whitney class, $\int_{S^2}w_2$, along an $S^2$ that surrounds it.

We can realize both of these as symmetry operators for 6D $\mathcal{N} = (1,0)$ SCFTs considered in \cite{Blum:1997fw, Blum:1997mm} (for a recent review see \cite{DelZotto:2023ahf}) that arise in the low-energy limit of $(Spin(32)/\mathbb{Z}_2)$-heterotic / Type I small instantons probing an ADE singularity. A key property of these SCFTs is that they possess a $Spin(32)/\mathbb{Z}_2$ flavor symmetry. Wrapping the 7-brane or 6-brane on the entire asymptotic boundary $S^3/\Gamma_{ADE}$ leads to a $\mathbb{Z}_2$-valued 0-form symmetry (this is the $\mathbb{Z}_2$ flavor center symmetry operator) and $\mathbb{Z}_2$-valued 1-form symmetry operator respectively.\footnote{Notice that when a 6-brane wraps the entire asymptotic boundary, we engineer a \textit{codimension-2} topological operator in the worldvolume of the 6D SCFT which is why it is a 1-form symmetry.} Backgrounds for these 0- and 1-form symmetries are simply associated with non-trivial Wilson line and $w_2$ for the $Spin(32)/\mathbb{Z}_2$ flavor backgrounds in the dual field theory.\footnote{6D SCFTs with non-trivial $w_2$ for the flavor bundle along compact directions has recently led to the construction of new 4D $\mathcal{N}=2$ SCFTs \cite{Ohmori:2018ona, Heckman:2022suy}, see also \cite{Dierigl:2020myk}.}

\section*{Acknowledgments}

We thank A. Debray, M. H\"ubner and X. Yu for several helpful discussions and collaboration on
related work. We thank H.T. Lam and S.-H. Shao for helpful discussions.
MD and MM thank UPenn and the organizers of the Strings and
Geometry '23 conference for hospitality during this work. The work of MD is
supported by the German-Israeli Project Cooperation (DIP) on \textquotedblleft
Holography and the Swampland\textquotedblright. The work of JJH is supported
by a University Research Foundation Grant at the University of Pennsylvania.
The work of JJH and ET is supported by DOE (HEP) Award DE-SC0013528.
The work of MM is supported by the Atraccion del Talento
Fellowship 2022-T1/TIC-23956 from Comunidad de Madrid.


\bibliographystyle{utphys}
\bibliography{R7BranesII}

\end{document}